# Spatial variations of the superconductor gap structure in $MgB_2$/Al composite

A. Sharoni, O. Millo[*], G. Leitus[+], S. Reich[+]

Racah Institute of Physics, The Hebrew University, Jerusalem 91904, Israel
[+] Dept. of Materials and Interfaces, The Weizmann Institute of Science, Rehovot 76100, Israel

Composites consisting of $MgB_2$ and Al, 11% by volume, undergo a transition to superconductivity onset at $T_C = 38$ K, close to the value reported for pure $MgB_2$. The transition appears to have broadened, as determined by both magnetic and transport measurements, possible due to the proximity effect and disorder. Spatially resolved tunneling spectroscopy at 4.2 K exhibit a distribution of gap structures, from BCS-like spectra with $2\Delta/k_B T_C = 3.2$ to spectra that are typical for proximity superconductors.

[*] e-mail: milode@vms.huj.ac.il

The recent discovery [1] of superconductivity around $T_C \sim 39$ K in the simple intermetallic compound $MgB_2$ is interesting for many reasons. This $T_C$ is much higher than the highest $T_C$ values reported for any non-oxide and non $C_{60}$-based material. Moreover, this transition temperature is above the limit predicted theoretically for conventional BCS superconductivity [2]. Nevertheless, the boron isotope effect measured by Bud`ko et al. [3] suggests that $MgB_2$ is a BCS phonon-mediated superconductor, where the high transition temperature is partly due to the low mass of boron that yields high phonon frequencies. Recent tunneling spectroscopy experiments performed by Sharoni et al. [4] and by Karapetrov et al.[5] also indicate that $MgB_2$ is a BCS-like *s*-wave superconductor. However, the superconductor gap structures observed in these two experiments look quite different, possibly due to different surface conditions or sample preparation procedure.

The ceramic nature of $MgB_2$ suggests that a composite with a malleable metal could render improved mechanical properties. In view of the relatively low melting and decomposition temperatures of $MgB_2$ ($T_M = 1,073$ K and decomposition above this temperature), the metal of choice should have a low melting temperature to yield sintering at about 600 K. Aluminum ($T_M = 933$ K), a malleable and a good conductor, offers the desirable physical properties. It should be stressed that for a sintering process performed in the range 500 to 700 K, there is no substitution of Mg by Al. Such a substitution was found to deteriorate the superconducting properties of $MgB_2$ [6]. The $MgB_2$/Al composite also provides a model system for the study of the roles played by the proximity effect and disorder in governing the global and local superconductor properties of $MgB_2$ (aluminum is a superconductor with low $T_C$, ~ 1.2 K). In this paper we present a study of the spatial variation of the superconductor gap structure of the $MgB_2$/Al composite using scanning tunneling microscopy (STM), in correlation with the global magnetic and transport properties of this system. The results are compared with previous data obtained on nominally pure $MgB_2$ samples, focusing on the effects of proximity and disorder.

Tunneling spectroscopy is widely used for characterizing the electronic properties of superconductors. In particular, scanning tunneling spectroscopy, enabled by the STM, is highly effective for studies of non-homogeneous superconductor systems due to the spatially resolved information it provides [7, 8]. The dI/dV vs V tunneling spectra yield direct information on the local quasi-particle density of states (DOS), and consequently on the superconductor gap structure [9]. For a disordered *s*-wave superconductor, where the quasi-particles experience significant scattering, the three main parameters that characterize the tunneling spectra are: 1. The gap parameter, $\Delta$ (or the energy gap $2\Delta$). 2. The quasi-particle life-time broadening parameter, $\Gamma$. 3. The zero bias conductance (ZBC) normalized to the normal tunnel-junction conductance (i.e., at a bias larger than $\Delta$). The first two parameters can be extracted by fitting the spectra to the Dynes function that describes tunneling from a normal electrode into a superconductor that exhibits a life-time broadened BCS DOS [10]. The picture becomes more involved in proximity systems consisting of normal and superconductor phases in good electrical contact. In the vicinity of a normal-superconductor interface the gap parameter reduces at the superconductor side, while a superconductor-like gap is induced at the normal side of the interface, sometimes referred to as a `mini-gap` [11]. The DOS at the normal side is best described (in the dirty limit) by the Usadel formalism [11, 12], but the tunneling spectra can nevertheless be well fit to the Dynes function [13].

We produced pellets 3 mm in diameter and 0.5 mm in height by compacting, under a pressure of $10^4$ Kg/cm$^2$, $MgB_2$ aluminum being 11% by volume, to produce a sintered material. Both the $MgB_2$ and the Al powders comprise of particles 1-3 µm in diameter. The sintering procedure was performed in vacuum at $T_S = 590$ K $= 0.63 T_M$, where $T_M = 933$ K is the melting temperature of Al.

Transport measurements were performed for few values of magnetic field up to 1 Tesla using the Van der Pauw method with four point silver paste contacts positioned on the rim of the pellets. The resistivity as a function of temperature is presented in Fig. 1, displaying a transition to superconductivity onset at $T_C = 38$ K at zero field and at H = 25 Oe. At 1 Tesla the onset is shifted down (as expected) to 35 K.

In Fig. 2 we plot field cooled (FC) and zero field cooled (ZFC) temperature-dependent magnetisation measured at 25 Oe with a SQUID $MPMS_2$ magnetometer. We observe an onset of the magnetic transition at 37.5 K. The superconducting volume fraction at 2 K, derived from the FC run, is 33 %. From the above transport and magnetic measurements we conclude that the sintering procedure in the mixture of $MgB_2$ and fine mesh Al powders did not significantly deteriorate the `global` superconducting properties of the $MgB_2$ particles. The main effect of Al in the $MgB_2$/Al composite is in broadening the transition, as compared to our previous measurements on nominally pure $MgB_2$ samples [4]. This broadening, probably due to the proximity effect and increased disorder, manifests itself also in the STM measurements, as discussed below.

For the STM measurements, the samples were either polished using diamond lapping-compound down to 0.25 µm, or were left as grown (the surfaces in the latter case were smooth enough to allow STM measurements). We found no significant difference between the results obtained for the polished and the unpolished samples. All samples were cleaned with dry nitrogen just before mounting in our homemade cryogenic STM and evacuating the sample space. The STM was immersed in liquid He and the sample and scan-head were cooled down to 4.2 K via He exchange gas. The tunneling dI-dV vs. V spectra were measured either directly using lock-in method or obtained by numerically differentiating the I-V curves measured simultaneously, both yielding similar results.

The tunneling characteristics exhibited spatial variations in the gap size, the ZBC and broadening. In general, three types of gap structures were observed in the tunneling spectra. In the first, the shapes conform well with the BCS DOS, with relatively small ZBC values, around 20% of the normal-state junction conductance, and relatively small broadening. These spectra were probably measured on the $MgB_2$ regions, far enough from large Al islands. In Fig. 3(a) we plot such a tunneling characteristic showing the maximal measured gap in these regions. Fit to the Dynes function (dashed line) yields a gap value of 5.2 meV and $\Gamma \sim 0.2\Delta$. With the measured $T_C = 38$ K we obtain that the ratio $2\Delta/k_BT_C$ approaches a value of 3.2, somewhat smaller than the theoretical value for a weak-coupling BCS superconductor (3.53). In some locations the dI/dV-V curves portrayed a significant contribution of in-gap states, suggesting enhanced quasi-particle excitation, yet still maintaining low ZBC and relatively sharp gap-edge shoulders. This phenomenon clearly manifests itself in fig. 3(b), where the increased in-gap spectral weight results in a nearly V-shaped spectrum that cannot be well reproduced by the Dynes function for any value of $\Gamma$.

The third type of tunneling spectra exhibit significantly larger normalized ZBC (around 50%) and broadening. Two representative curves of this type are presented in Fig. 4. It is evident that the gap edge shoulders, among the most significant signatures of a BCS quasi-particle DOS, are highly smeared (as compared to those observed in the data presented in Fig. 3). In fact, these spectra resemble the `mini-gap` structures induced via the proximity effect in disordered normal metals that are in contact with a superconductor [11]. It is therefore highly possible that such spectra were measured on the Al regions, (we recall that Al is a superconductor with $T_C$ smaller than 4.2 K). Unfortunately, in contrast to our previous studies of the proximity effect [7, 13], we could not identify in our topographic images any clear boundaries between different regions, thus we cannot unambiguously confirm this picture. The gap parameters extracted for these type of spectra from fitting to the Dynes function were in the range $\Delta = 3.5\text{-}4.5$ meV. However, due to the large broadening parameters needed to obtain good fits, $\Gamma = 0.5\Delta\text{-}0.6\Delta$, there is a large uncertainty in these gap values. In addition to gap structures, we have also measured Ohmic I-V characteristics, displaying normal-metallic behavior. Such curves were probably measured well within Al regions, far enough (more than the coherence length in the normal Al region) from interfaces with $MgB_2$.

It is instructive to compare the results for the $MgB_2$/Al composite presented here with previous measurements performed on nominally pure $MgB_2$ samples by Sharoni et al. [4] and by Karapetrov et al. [5]. The samples studied by Sharoni et al. were prepared by heating a stoichiometric ratio of Mg and B elements to 950 C in a sealed Ta capsule. The tunneling spectra measured on these samples exhibited larger gaps, between 5 to 7 meV ($2\Delta/k_BT_C$ varied between 3 to 4.2), smaller broadening, typically $\Gamma \sim 0.1\Delta$, and smaller ZBC (that vanished nearly everywhere). Moreover, V-shaped tunneling spectra such as shown in Fig. 3(b) were *not* observed in this previous study [4]. These differences in the local spectral properties are consistent with the global behavior exhibited by these two systems. The samples studied in Ref. [4] have shown a sharper magnetic transition onset at a slightly higher $T_C$ (39 K), as compared to the $MgB_2$/Al composite. The broader transition reported here also suggests the existence of large areas having degraded local superconductivity, due both to the proximity effect and to disorder. The tunneling spectra presented in Ref. [5] closely resemble, in shape, broadening and in their ZBC values, those shown in Fig. 4 (that we attribute to proximity-induced `mini-gaps` in Al). However, the gap parameters reported by Karapetrov et al., ~ 5.2 meV with small spatial variations, were close to the values we measured on areas displaying the highest quality superconductivity in the $MgB_2$/Al composite (Fig. 3). The variance in the energy gap values and gap structures observed in measurements performed on different $MgB_2$ samples prepared in different ways emphasizes the need for performing measurements on pure single-crystal materials. There, one can look for possible directional dependence of the tunneling spectra, which may shed light on the origin of these sample to sample variations and, more important, on the coupling mechanism in this system. Indeed, recent measurements performed on etched $MgB_2$ pellets and *c*-axis oriented $MgB_2$ films exhibited directional-dependent variations of the gap structure and magnitude, attributed to an anisotropic *s*-wave order parameter [14].

In summary, while the onset of superconductivity in our $MgB_2$/Al composite ($T_C$ = 38 K) was close to that observed for pure $MgB_2$ ($T_C$ = 39.5), the transition was found to be broader, probably due to the proximity effect and disorder induced by the introduction of Al. This `global` sample behavior manifested itself also in a wide spatial distribution of gap parameters and shapes of the tunneling spectra. These spectra revealed `mini-gap` structures, typical for disordered proximity superconductors, enhanced in-gap quasi-particle excitations, as well as clear BCS-like gap structures. The maximal observed gap parameter was 5.2 meV, yielding $2\Delta/k_BT_C$ = 3.2, consistent with BCS superconductivity.

This work was supported by the Israel Science Foundation of the Israel Academy of Science and Humanities.

**Figures and Captions**

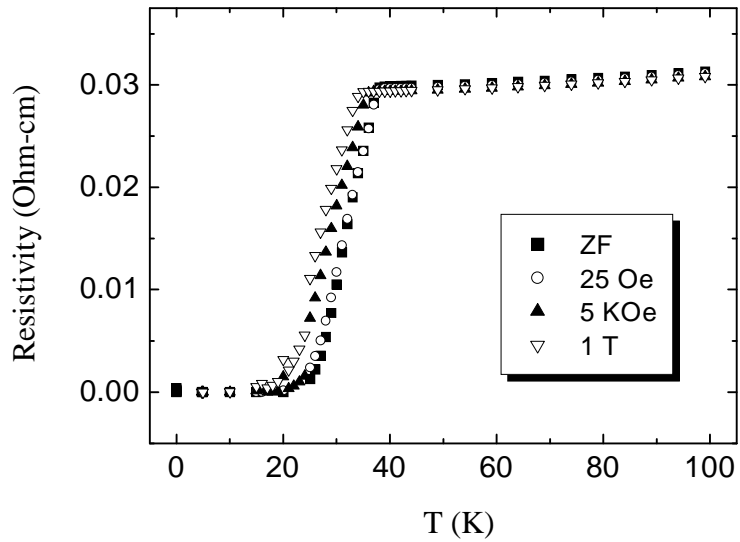

**Fig. 1:** Resistivity vs. temperature of $MgB_2$/Al composite at different magnetic fields. Zero field (ZF) measurements show a transition to superconductivity onset at 38 K, while at 1 Tesla the onset is shifted down to 35 K.

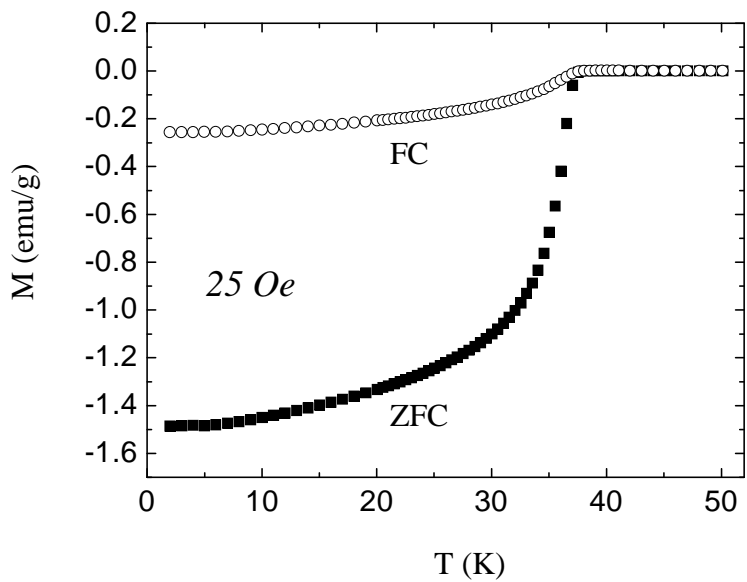

**Fig. 2:** Field cooled (FC) and zero field cooled (ZFC) temperature-dependent magnetisation at 25 Oe showing a transition to superconductivity onset at 37.5 K.

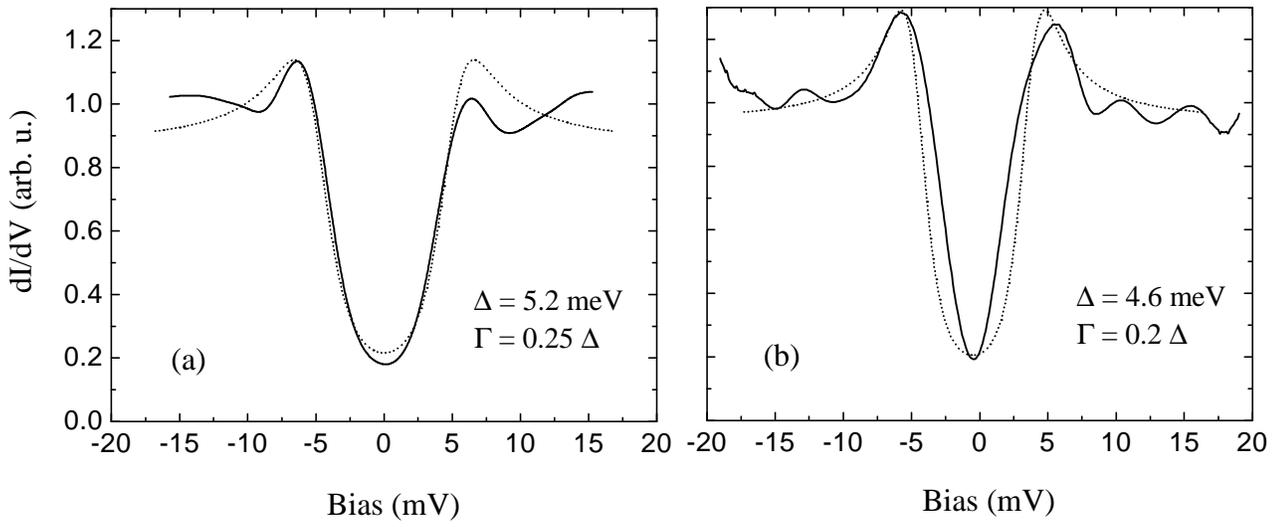

**Fig. 3:** Tunneling characteristics acquired on regions exhibiting BCS-like behavior with low zero bias conductance (solid curves). Curve (a) shows the maximal gap observed in these regions while curve (b) exhibits enhanced contribution of in-gap states. The dotted lines represent fits to the Dynes function with parameters denoted in the figure.

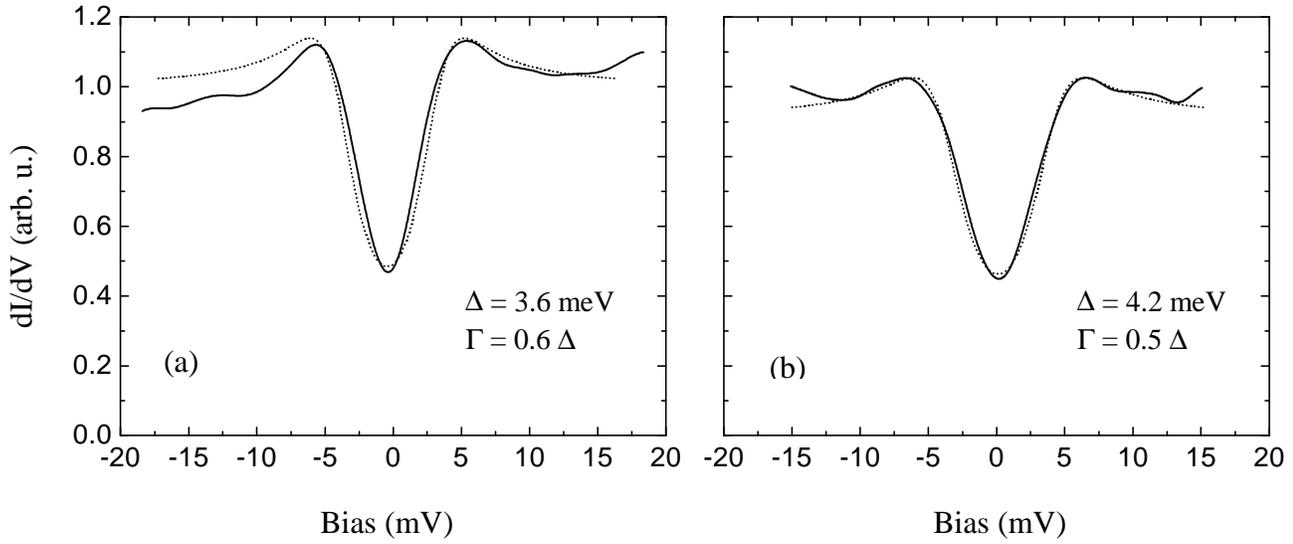

**Fig. 4:** Tunneling characteristics acquired on regions exhibiting degraded BCS behavior (solid curves). The dotted lines represent fits to the Dynes function with parameters denoted in the figure.